\documentclass[twocolumn,epjc3]{svjour3}  
\smartqed  % flush right qed marks, e.g. at end of proof
\RequirePackage{graphicx}

\usepackage{amsmath}
\usepackage{txfonts}
\usepackage{balance}
\journalname{Eur. Phys. J. C}

\begin{document}

\title{Constraints on dark matter annihilation from M87%\thanksref{t1}
}
\subtitle{Signatures of  prompt and inverse-Compton gamma rays}


\author{Sheetal Saxena\thanksref{e1,addr1}
        \and
        Alexander Summa\thanksref{addr1} %etc.
\and Dominik Els\"asser\thanksref{addr1} %etc.
\and Michael R\"uger\thanksref{addr1} %etc.
\and Karl Mannheim\thanksref{addr1} %etc.
}

\thankstext{e1}{e-mail: saxena@astro.uni-wuerzburg.de}


\institute{Institute for Theoretical Physics and Astrophysics, University of W\"urzburg, Campus Hubland Nord, Emil-Fischer-Str. 31, 97074 W\"urzburg, Germany \label{addr1}
         }
\date{Received: date / Accepted: date}

\maketitle

\begin{abstract}

As the largest mass concentrations in the local Universe, nearby clusters of galaxies and their central galaxies are prime targets in searching for indirect signatures of dark matter annihilation (DMA). We seek to constrain the dark matter annihilation emission component from multi-frequency observations of
the central galaxy of the Virgo cluster. The annihilation emission component is modeled by the prompt 
and inverse-Compton gamma rays from the hadronization of annihilation products from
generic weakly interacting dark matter particles.  This component is fitted to
the excess of the observed data above the spectral energy distribution (SED) of the jet in M87,
described with a best-fit synchrotron-self-Compton (SSC) spectrum. While this result is not sufficiently significant  
to claim a detection, we emphasize that a dark matter ``double hump signature'' can be used
to unambiguously discriminate the dark matter emission component from the variable 
jet-related emission of M87 in future, more extended observation campaigns.

\end{abstract}

\section{Introduction}
\label{sec:1}
The thermal freeze-out of weakly interacting particles (so-called 'WIMPs') 
with masses at the electroweak symmetry breaking scale $E_{\rm ew}=
1/\sqrt{2^{1/2}G_{\rm F}}\approx 246$~GeV
leads to a relic density agreeing with the observed one $\Omega_{\rm dm}h^2=0.1123\pm0.0035$ \cite{Lee1977,Jarosik2011}.
This so-called WIMP miracle provides a natural solution to the dark matter problem.
In the present-day Universe,  WIMPs can still annihilate in regions with large mass concentrations.
The Galactic Center \cite{Aharonian2006,Albert2006}, dwarf galaxies in the halo of the Milky Way \cite{Aliu2009,Albert2008,Aleksi'c2011,Aharonian2008,Acciari2010}, and galaxy clusters \cite{Aleksi'c2010,Ackermann2010} have been studied to obtain limits on
dark matter annihilation emission.
Annihilation gives rise to pairs of heavy quarks, leptons or vector bosons, which decay leading
to the emission of gamma rays, electrons, positrons, and neutrinos.  The secondary radiation shows
a peak at $E_{\rm ew}/20\approx 10$~GeV reflecting the high multiplicity of secondary particles
in a typical annihilation event.
The canonical peak energy lies at the high end of
the bandwidth of the Fermi-LAT detector, and at the low end of the accessible energy
range for ground-based
Cherenkov telescopes. A very well motivated WIMP candidate is the Lightest Supersymmetric Particle (LSP): the neutralino \cite{Bertone2005}. Annihilating neutralinos produce vector bosons, leptons or quarks. The differential gamma-ray flux \cite{Evans2004} produced through their subsequent hadronization and decay from a source at distance $D$ and volume $V$ is given by:
\begin{equation}
\left(\frac{\mathrm{d}\Phi}{\mathrm{d}E}\right)_{\pi^{0}}=\frac{1}{4\pi}\frac{f\left\langle\sigma_{A}\,\nu\right\rangle}{2m_{\chi}^{2}}\frac{\mathrm{d}N_{\gamma}}{\mathrm{d}E}\frac{1}{{D}^{2}}\int\limits_{\rm{M87}} \!  \mathrm{d}V\,\rho_{\rm NFW}^2
\end{equation}
where $f$ is the boost factor that accounts for enhancement due to sub-halo clumping of dark matter in the M87 halo, $\left\langle \sigma_{A}\nu\right\rangle$ is the thermally averaged annihilation cross section, $m_{\chi}$ is the mass of the dark matter particle, and $\mathrm{d}N_{\gamma}/\mathrm{d}E$ is the gamma photon spectrum coming from the decay of neutral pions from hadronization in the annihilation process (prompt pion emission) \cite{Saxena2011}. $\rho$ is taken to be $\rho_{\rm NFW}=\rho_{\rm NFW}(r)$ the Navarro-Frenk-White dark matter density profile \cite{Navarro1996} of the dark matter halo obtained from numerical simulations.

A minimum mass scale of $10^{-5}M_\odot$ for the sub-halo clum\-ping
has been inferred from the transfer function of density perturbations in the early Universe for weakly interacting dark matter particles \cite{Green2004}. 
However, tidal interactions with the baryon-dominated cores of dark matter halos and supernova feedback could destroy these structures to a  large extent.   
Therefore, the boost factor $f=\int \rho^2 \rm{d}V/\int\rho_{\rm NFW}^2\rm{d}V$ is introduced as a free parameter to account for the unknown enhancement due to sub-halo clumping.

The inverse-Compton interaction is the up-scattering of photons by high-energy charged particles. The differential gamma-ray flux given by Eq. (2) describes inverse-Compton scattering off the cosmic microwave background by relativistic electrons and positrons. 
$b\left(E'\right)$ is the total rate of electron/positron energy loss due to inverse-Compton scattering as in \cite{Cirelli2009}, $P\left(E,E'\right)$ is the differential power emitted into photons of energy $E$ by an electron/positron with energy $E'$, and $\mathrm{d}N_{e}/\mathrm{d}\widetilde{E}$ is the spectrum of secondary electrons and positrons \cite{Saxena2011}: 
\begin{eqnarray}
\left(\frac{\mathrm{d}\Phi}{\mathrm{d}E}\right)_{\rm{IC}}&=& \frac{1}{E}\frac{f\left\langle\sigma_{A}\,\nu\right\rangle}{4\pi m_{\chi}^{2}}\frac{1}{{D}^{2}}\int\limits_{\rm{M87}} \!  \mathrm{d}V\,\rho_{\rm NFW}^2\left(r\right)\nonumber \\
&&\times\int\limits_{m_{e}}^{m_{\chi}}\mathrm{d}E'\,\frac{P\left(E,E'\right)}{b\left(E'\right)}\int\limits_{E'}^{m_{\chi}}\mathrm{d}\widetilde{E}\,\frac{\mathrm{d}N_{e}}{\mathrm{d}\widetilde{E}}
\end{eqnarray}

The main challenge in constraining the putative dark matter annihilation component is to discriminate it
against gam\-ma rays from astrophysical sources and cosmic ray interactions.
Here, we show as an exemplary case study 
how recent multi-frequency data of the center of the Virgo cluster, harboring the giant cD galaxy M87 with its
gamma-ray emitting radio jet,  can be used to constrain dark matter particles.  Note that we do not refer to the inverse-Compton scattering in so-called leptophilic dark matter annihilation scenarios, but include 
the inverse-Compton emission component due to  the electrons and positrons
from the decay of isospin-symmetric annihilation products.

We adopt a  distance of $(16.5\pm1)$~Mpc \cite{NED} to M87 and therefore omit redshift corrections to the energy throughout the paper.  Section 2 describes the data sets chosen for the
study.  The radiation code employed to model the spectral energy distribution of M87 is explained
in Section 3, and physical parameters inferred from the fit of the data are briefly discussed to show
that the model is a viable interpretation of the nonthermal particle content in the jet of M87.
The generic model for the weakly interacting massive particles
and their radiative signatures is described in Section 4, and a fit of a component from dark matter annihilation to the excess above the astrophysical model is presented.  
Finally, we discuss the results and draw conclusions regarding future search strategies.

\section{Data}
\label{sec:2}
The dataset used here is comprised of observations of M87 by the Chandra X-ray Observatory, 
Fermi-LAT, MAGIC, and the H.E.S.S. system of Cherenkov telescopes \cite{Harris2009,Abdo2009,Berger2011,Aharonian2006a}. As presented in \cite{Abdo2009}, additional 
observational data exist
for the radio-to-optical regime. However, we do not use these to constrain the fit, but rather only require the projected
emission to not exceed these observations, since in this wavelength regime the unavoidable contamination due to starlight and dust from
the central region of M87 is very difficult to assess.  There may also be hidden nonthermal components
in this energy region unrelated to the emission region where the high-energy emission originates from.

The data sample was not taken contemporaneously, but we carefully checked to avoid inclusion of an observation containing a significant 
flare. All the data used here can thus be considered a representative long-term average, reasonably well describing the steady and low-state spectral energy distribution of the high-energy emission component in M87.

%\cite{RefB} and \cite{RefJ}.

\section{Synchrotron-self-Compton emission}
\label{sec:3}
The nonthermal emission from the relativistic jet emerging from the nucleus of M87
follows a spectral energy distribution across almost 20 orders of magnitude which
can be described by the synchrotron and self-Compton radiation produced by
electrons (and positrons) accelerated in the jet, presumably by shock waves.
Whereas the high amplitude flares track single shock waves or magnetic reconnection
events in the expanding flow, guided by the helical structure of the
magnetic field, the steady-state (long-term average)
emission which dominates the energy output from the jet of M87 corresponds
to the superposition of a large number of shock waves \cite{Blandford1979}, a strong stationary shock such as the reconfinement shock \cite{Marscher2008} or
the emission from a sheath surrounding the spine of the jet \cite{Tavecchio2004}. 

To obtain a fit for the broad-band spectral energy distribution of
M87, we used an implementation of the SSC model \cite{Ruger2010} in which
the cooling break in the electron spectrum is self-consistently determined,
and the cross section for Compton scattering in the Klein-Nishina regime is
accurately treated.  We also took into account the inhomogeneous nature
of the emission zone, by considering only the emission from the inner jet
for the fit, i.e. we considered the radio to optical emission, related to the
emission from larger scales in the jet, as an upper bound
for the model.

Adopting for the Doppler factor $\delta=3.9$ (bulk Lorentz factor $\Gamma=2.3$) as
in the model fitted to the data by \cite{Abdo2009}
we obtain a fit with $\chi^2/\nu=2.5$. 
Physical parameters of the emission zone are the source radius $R_b=3.5\times 10^{13}$ cm and the magnetic
field strength $B=3$ G. For the injected electron power law distribution with
exponential cut-off, we obtain a differential slope of $s=2.2$, maximum Lorentz factor
$\Gamma_{max}=10^8$ and normalization factor $K=10^{6}\,\rm{cm^{-3}s^{-1}}$ (cf. \cite{Ruger2010}). This results in an
injection luminosity of $L_{inj}=3\times10^{41}$~erg~s$^{-1}$ consistent with the energetics of the 
jet inferred from its large scale radio structure \cite{Hardee1982}.  
Model fits including the low-energy continuum \cite{Abdo2009} yield
different results, but fail to produce an acceptable fit including the very high energy observations.

For the black hole mass of M87
$M_{\rm BH}= 6.4\times10^9M_{\odot}$, the Eddington luminosity is given by $L_{\rm E}= 8.32\times
10^{47} $ erg~s$^{-1}$. Thus, the nonthermal power release amounts to only 
$\sim 10^{-6}L_{\rm E}$.  Although a sub-Eddington state of the black hole is generally
expected for high-peaked blazars, the extremely low power is peculiar.
In fact,  it is not possible to model the data by simply adjusting the
Doppler factor of an  SSC fit obtained for high-peaked blazars, i.e. by changing the inclination
of the jet axis with respect to the observer. The idea of treating M87 as a misaligned blazar
\cite{Mucke2001} does not seem to be sufficient. Furthermore the
large difference between break energy and cut-off is not typical for blazar
spectra.  This might be related to the extremely low accretion rate, indicating
that the AGN is fading out due to a lack of accretable matter, or to the fact that the
observed emission is dominated by the emission from a low-bulk-Lorentz-factor sheath surrounding the jet.

The SSC fit of the data obtained in this way provides an accurate and practically unique model for the SED, and the small size of the emission region is in line with observations of short-time variability.

\begin{figure}
  \includegraphics[width=0.45\textwidth]{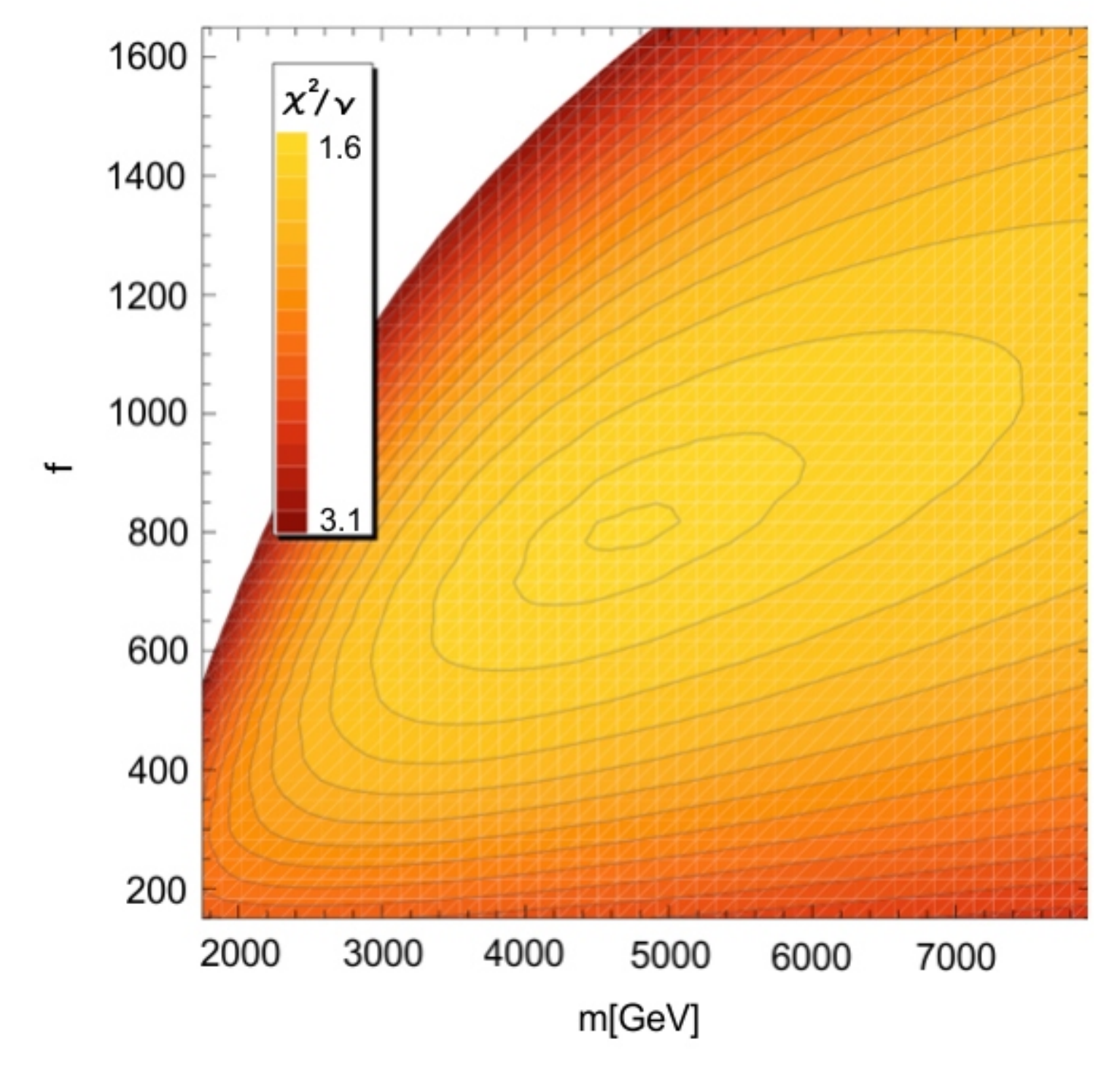}
\caption{Result of the combined fit  including the spectral energy distributions due to the SSC mechanism in the jet of M87  and the annihilation of  dark matter particles.
The boost factor $f$ accounts for sub-halo clumping, and $m$ denotes the generic WIMP mass.}
\label{fig:1}       % Give a unique label
\end{figure}

\section{Fitting the excess emission component with dark matter annihilation}
\label{sec:4}

\begin{figure*}
  \includegraphics[width=0.95\textwidth]{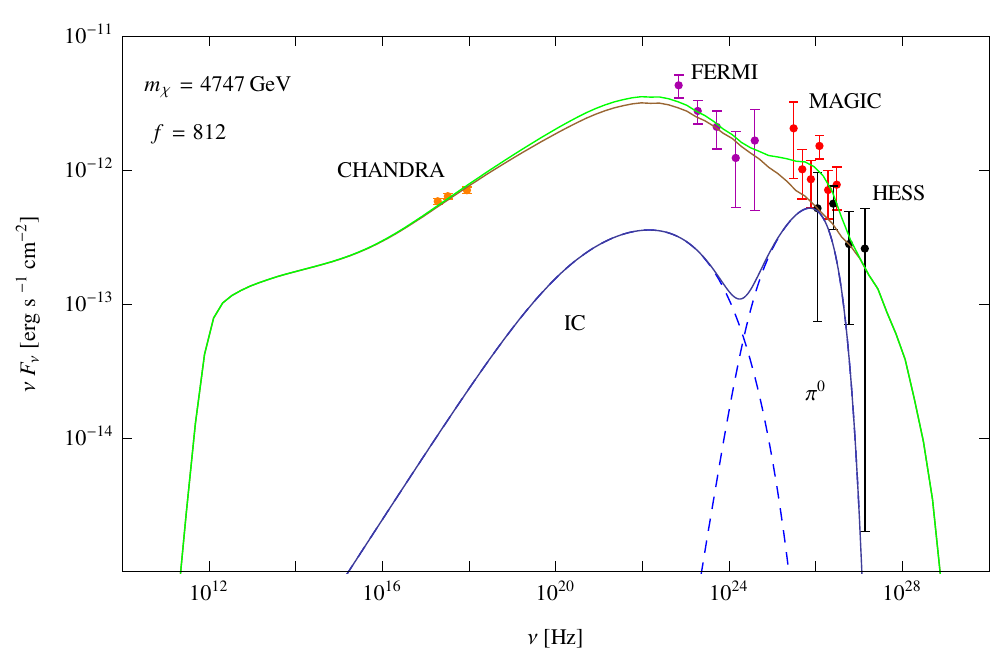}
\caption{Spectral energy distribution of the best-fit combined SSC model for M87 and dark matter model, including all data points used in this analysis \cite{Harris2009,Abdo2009,Berger2011,Aharonian2006a}.
The inset legend provides the mass and boost factor for the generic WIMP with cross section $\left\langle \sigma_{A}\nu\right\rangle=3\times10^{-24}\,\rm{cm}^3\rm{s}^{-1}$.}
\label{fig:2}       % Give a unique label
\end{figure*}

To now fit an additional component due to dark matter annihilation, and study whether the statistical agreement
between model and observations can thereby be further improved, we assume a generic species of annihilating WIMPs
with rest mass in the GeV-TeV range and a thermally averaged annihilation cross section of $\left\langle \sigma_{A}\nu\right\rangle=3\times10^{-24}\,\rm{cm}^3\rm{s}^{-1}$. It has been shown that many models with multi-TeV masses provide large pair annihilation cross sections which are still in agreement with the thermal freeze-out of these particles in the early Universe \cite{Profumo2005}. The choice of the annihilation cross section also reflects the possible existence of boost factors in the particle physics sector and should be considered as moderate upper limit (cf. \cite{Pinzke2011}). For the dark matter distribution of M87, we use the analysis of \cite{McLaughlin1999}, resulting in a description of the halo according to the Navarro-Frenk-White \cite{Navarro1996} model.
The normalization of the intensity of the resulting
emission is furthermore fixed by choosing the boost factor from unresolved substructure in the halo of M87, which from
recent numerical experiments \cite{Springel2008} is expected to be of order $f=10^{2}-10^{3}$. The spectra for the emission due to the decay of charged and neutral pions from hadronization in the annihilation process are generated using the DarkSUSY code \cite{Gondolo2004}.
A total of $10^6$ realizations of neutralinos are produced to obtain the average secondary spectra for generic WIMPs motivated by supersymmetric theories.

Here we also include a treatment of the inevitable contribution~of inverse-Compton
emission from energetic electrons/positrons from the decay of the charged pions up-scat\-ter\-ing cosmic microwave background photons. 
By using the known number density of these 2.7 K background photons 
($413$~cm$^{-3}$), the respective contribution
from inverse-Comp\-ton emission can be of the same order as the prompt pion decay emission. This results in a telltale ``double hump structure'' of the dark matter related emission.
Employing a $\chi^{2}$-test, we search for the best-fit model of the active galactic nucleus (AGN) as discussed in the previous section, and the complete dark matter related model. The results of the fit are shown in Fig. 1.

A minimal $\chi^{2}$ value of 1.6 is achieved for a particle mass of 4.7 TeV and a boost factor of 812, both well within the range of
values accessible in state-of-the-art particle physics and numerical models, as discussed in the previous section. %While the reduced $\chi^{2}$ is still well above unity,
Assuming that the main multi-wavelength features of such a generic dark matter model can be described by two additional parameters (neutralino mass and boost factor) compared to the SSC model alone, it can be shown by nonlinear regression analysis that the deviation of the data points with respect to the SSC model amounts to $2\sigma$. We note that this result is not significant enough to claim the realization of a certain dark matter model, but even our simplified assumptions show the capability of multi-wavelength methods for unvealing the characteristics of the constituents of dark matter.
In Fig. 2, we show the resulting spectral energy distribution from the combined SSC-DMA model.

\section{Discussion and conclusions}
\label{sec:5}

We show that while synchrotron-self-Compton emission can fit the observations reasonably well, the introduction of a component due to annihilating dark matter particles margi\-nal\-ly improves the agreement between theoretical modeling and multi-frequency observations of the long-term average SED of the radio galaxy M87.  It is essential to include the inverse-Compton emission component in the dark matter annihilation model.
The method thus shows great sensitivity in ruling out dark matter models at the TeV scale which is
innate to some supersymmetric extensions 
of the Standard Model, while annihilation cross section and additional boost from substructures are concordant with the paradigm of thermal freeze-out of such a particle at the onset of hierarchical structure formation.
Although the significance of this result is not  sufficient to claim evidence of a specific particle, it is
encouraging further studies. 

Since the SSC component generally shows variability while the DMA component remains steady, we emphasize that the significance of the marginal excess reported here would 
increase during low states of the SSC emission. Therefore we encourage to extend the temporal coverage of 
multi-wavelength observations of M87. Measuring the putative \linebreak dark matter ``double hump structure`` in the spectrum with a high significance would open up the possibility to extract the dark matter
properties (WIMP mass, annihilation cross section and boosting due to substructure) more accurately.

The high mass scale of the generic WIMP favored in our model here is in-line with recent findings at the LHC \cite{Baer2009}. 
A high mass scale may also help in explaining the measured extragalactic gamma-ray background (EGB) \cite{Elsasser2005},
after a careful reassessment of the Fermi-measured EGB \cite{Abdo2010}. Detection of high-mass WIMPs by
elastic scattering experiments \cite{CDMSIICollaboration2010} will be difficult due to the suppression of the event rate by the low number density of
WIMPs in the Galactic halo.

\bibliographystyle{springer}
\balance
\bibliography{M87}

\begin{thebibliography}{10}

\bibitem{Lee1977}
B.~W. Lee, S.~Weinberg, Phys. Rev. Lett. \textbf{38}, 1237 (1977)

\bibitem{Jarosik2011}
N.~{Jarosik}, et~al., \apjs \textbf{192}, 14 (2011)

\bibitem{Aharonian2006}
F.~{Aharonian}, et~al., \nat \textbf{439}, 695 (2006)

\bibitem{Albert2006}
J.~{Albert}, et~al., \apjl \textbf{638}, L101 (2006)

\bibitem{Aliu2009}
E.~{Aliu}, et~al., \apj \textbf{697}, 1299 (2009)

\bibitem{Albert2008}
J.~{Albert}, et~al., \apj \textbf{679}, 428 (2008)

\bibitem{Aleksi'c2011}
J.~{Aleksi{\'c}}, et~al., \jcap \textbf{6}, 35 (2011)

\bibitem{Aharonian2008}
F.~{Aharonian}, et~al., Astroparticle Physics \textbf{29}, 55 (2008)

\bibitem{Acciari2010}
V.~A. {Acciari}, et~al., \apj \textbf{720}, 1174 (2010)

\bibitem{Aleksi'c2010}
J.~{Aleksi{\'c}}, et~al., \apj \textbf{710}, 634 (2010)

\bibitem{Ackermann2010}
M.~{Ackermann}, et~al., \apjl \textbf{717}, L71 (2010)

\bibitem{Bertone2005}
G.~{Bertone}, D.~{Hooper}, J.~{Silk}, \physrep \textbf{405}, 279 (2005)

\bibitem{Evans2004}
N.~W. {Evans}, F.~{Ferrer}, S.~{Sarkar}, \prd \textbf{69}, 123501 (2004)

\bibitem{Saxena2011}
S.~{Saxena}, D.~{Els{\"a}sser}, M.~{R{\"u}ger}, A.~{Summa}, K.~{Mannheim},
  Proceedings of the 32nd International Cosmic Ray Conference  (2011)

\bibitem{Navarro1996}
J.~F. {Navarro}, C.~S. {Frenk}, S.~D.~M. {White}, \apj \textbf{462}, 563 (1996)

\bibitem{Green2004}
A.~M. {Green}, S.~{Hofmann}, D.~J. {Schwarz}, \mnras \textbf{353}, L23 (2004)

\bibitem{Cirelli2009}
M.~{Cirelli}, P.~{Panci}, Nuclear Physics B \textbf{821}, 399 (2009)

\bibitem{NED}
NASA/IPAC Extragalactic Database, http://ned.ipac.caltech.edu

\bibitem{Harris2009}
D.~E. {Harris}, C.~C. {Cheung}, {\L}.~{Stawarz}, J.~A. {Biretta}, E.~S.
  {Perlman}, \apj \textbf{699}, 305 (2009)

\bibitem{Abdo2009}
A.~A. {Abdo}, et~al., \apj \textbf{707}, 55 (2009)

\bibitem{Berger2011}
K.~{Berger}, D.~{Dominis Prester}, F.~{Tavecchio}, T.~{Terzi{\'c}}, {for the
  MAGIC Collaboration}, arXiv:1109.5879v1  (2011)

\bibitem{Aharonian2006a}
F.~{Aharonian}, et~al., Science \textbf{314}, 1424 (2006)

\bibitem{Blandford1979}
R.~D. {Blandford}, A.~{Konigl}, \apj \textbf{232}, 34 (1979)

\bibitem{Marscher2008}
A.~P. {Marscher}, et~al., \nat \textbf{452}, 966 (2008)

\bibitem{Tavecchio2004}
F.~{Tavecchio}, et~al., \apj \textbf{614}, 64 (2004)

\bibitem{Ruger2010}
M.~{R{\"u}ger}, F.~{Spanier}, K.~{Mannheim}, \mnras \textbf{401}, 973 (2010)

\bibitem{Hardee1982}
P.~E. {Hardee}, \apj \textbf{261}, 457 (1982)

\bibitem{Mucke2001}
A.~{M{\"u}cke}, R.~J. {Protheroe}, Astroparticle Physics \textbf{15}, 121
  (2001)

\bibitem{Profumo2005}
S.~{Profumo}, \prd \textbf{72}, 103521 (2005)

\bibitem{Pinzke2011}
A.~{Pinzke}, C.~{Pfrommer}, L.~{Bergstrom}, arXiv:1105.3240v3  (2011)

\bibitem{McLaughlin1999}
D.~E. {McLaughlin}, \apjl \textbf{512}, L9 (1999)

\bibitem{Springel2008}
V.~{Springel}, et~al., \mnras \textbf{391}, 1685 (2008)

\bibitem{Gondolo2004}
P.~{Gondolo}, et~al., \jcap \textbf{7}, 8 (2004)

\bibitem{Baer2009}
H.~{Baer}, X.~{Tata}, in \emph{Physics at the Large Hadron Collider}, edited by
  {Datta, A., Mukhopadhyaya, B., Raychaudhuri, A., Gupta, A.~K., Khetrapal,
  C.~L., Padmanabhan, T., Raychaudhuri, A., \& Vijayan, M. }, Springer India
  (2009)

\bibitem{Elsasser2005}
D.~{Els{\"a}sser}, K.~{Mannheim}, Physical Review Letters \textbf{94}, 171302
  (2005)

\bibitem{Abdo2010}
A.~A. {Abdo}, et~al., Physical Review Letters \textbf{104}, 101101 (2010)

\bibitem{CDMSIICollaboration2010}
{CDMS II Collaboration}, Science \textbf{327}, 1619 (2010)

\end{thebibliography}

\end{document}